\documentclass[11pt,preprint]{aastex}
\begin{document}

\slugcomment{Accepted to the {\it Astrophysical Journal}}

\title{The Initial Mass Function of Low-Mass Stars and Brown Dwarfs in 
Taurus\altaffilmark{1}}

\author{K. L. Luhman}
\affil{Harvard-Smithsonian Center for Astrophysics, 60 Garden St., 
Cambridge, MA 02138}

\email{kluhman@cfa.harvard.edu}

\altaffiltext{1}{Visiting Astronomer, Kitt Peak National Observatory,
National Optical Astronomy Observatories, which is operated by the
Association of Universities for Research in Astronomy, Inc.\ (AURA) under
cooperative agreement with the National Science Foundation.}

\begin{abstract}

By combining deep optical imaging and infrared spectroscopy with data from the
Two-Micron All-Sky Survey (2MASS) and from previous studies 
(e.g., Brice\~{n}o et al.), I have measured the Initial Mass Function (IMF) 
for a reddening-limited sample in four fields in the Taurus star forming 
region.  This IMF is representative of the young populations within these
fields for masses above 0.02~$M_{\odot}$.
Relative to the similarly derived IMF for the Trapezium Cluster (Luhman et al.),
the IMF for Taurus exhibits a modest deficit of stars above one solar mass 
(i.e., steeper slope), the same turnover mass ($\sim0.8$~$M_{\odot}$), and 
a significant deficit of brown dwarfs. If the
IMF in Taurus were the same as that in the Trapezium, $12.8\pm1.8$ brown dwarfs 
($>0.02$~$M_{\odot}$) are expected in these Taurus fields where only 
one brown dwarf candidate is found.
These results are used to test theories of the IMF. 

\end{abstract}

\keywords{infrared: stars --- stars: evolution --- stars: formation --- stars:
low-mass, brown dwarfs --- stars: luminosity function, mass function ---
stars: pre-main sequence}

\section{Introduction}

A great deal of insight into the process of star formation can be gained by
measuring the initial mass function (IMF) of low-mass stars and brown dwarfs.
The presence of a flattening or turnover in the IMF, the value of such a 
characteristic mass, the shape of the IMF into the substellar regime, 
the minimum mass at which objects form in isolation, and the variation of 
these properties with environmental conditions can provide discriminating tests 
of the wide range of theories for how the masses of stars are determined.

In an effort to measure accurately and precisely the low-mass IMF under 
a wide range of conditions, the populations within the nearby star forming 
clusters IC~348 (Luhman et al.\ 1998b), $\rho$~Oph (Luhman \& Rieke 1999), 
and the Trapezium Cluster (Luhman et al.\ 2000) were studied in detail.
As summarized by Luhman et al.\ (2000), the stellar IMFs in these clusters were
found to be similar over a wide range of stellar densities
($\rho$~Oph, $n=0.2$-$1\times10^3$~pc$^{-3}$; IC~348, $n=1\times10^3$~pc$^{-3}$;
Trapezium, $n=1$-$5\times10^4$~pc$^{-3}$), while modest variations in the
substellar IMFs were not ruled out. The data for star forming clusters (see
also Hillenbrand \& Carpenter 2000), young open clusters (Bouvier 
et al.\ 1998; Barrado y Navascu\'{e}s et al. 2000), 
and the field (Reid et al.\ 1999) were
shown to be consistent with the same IMF, one that is flat or rises 
slowly from the substellar regime to about 0.6~$M_{\odot}$, and then 
rolls over into a power law that continues from about 1~$M_{\odot}$ to 
higher masses, consistent the notion that the majority of field stars 
were born in clusters (Lada, Strom, \& Myers 1993). 
Meanwhile, a different IMF characterizes Galactic globular
clusters (Paresce \& De Marchi 2000), one that can be described by 
a log-normal form with a characteristic mass of 0.33~$M_\odot$. 
These studies of star forming and open clusters, the field, and globular
clusters represent the most reliable measurements of the low-mass IMF to date. 
The observed variation in the IMF from the field and Galactic clusters
to globular clusters motivates
further studies of the low-mass IMF under a greater variety of environments.

Whereas the young stellar populations from the previous studies represent
the clustered mode of star formation ($n=10^2$-$10^4$~pc$^{-3}$),
the stars in the Taurus-Auriga molecular cloud are forming in relative 
isolation (1-10~pc$^{-3}$).
Because of the compact nature of star forming clusters, candidate members 
can be efficiently identified through optical and infrared (IR) imaging and 
confirmed through multi-object spectroscopy.
The completeness limits in mass and reddening of such samples
are easily defined with IR photometry, which is essential for 
a measurement of the IMF (see, e.g., Luhman et al.\ 2000). 
In contrast, the population in the Taurus star forming region 
is spread across a much larger area of sky, dramatically increasing the
field star contamination and reducing the observing efficiency.
Consequently, unlike the unbiased samples defined for clusters, young stars 
in Taurus are usually identified through specific selection criteria 
(e.g., emission in H$\alpha$, X-rays, or the far-IR). Such samples are 
not satisfactory for measurements of the IMF because the biases and 
mass completeness limits are not readily quantified.

In a few recent studies, representative measurements of the low-mass 
population in Taurus have been attempted. 
Strom \& Strom (1994) and Luhman \& Rieke (1998) (hereafter LR98)
used optical and IR imaging and spectroscopy to estimate the IMF in L1495, 
one of the more well-defined stellar aggregates in Taurus. 
To search for objects at lower masses over a larger area, 
Brice\~{n}o et al.\ (1998) (hereafter BHSM) obtained
$RI$ imaging of the L1495, B209, L1529, and L1551 dark clouds.
Because of the proximity and youth of Taurus, the locus of members in the BHSM
optical color-magnitude diagram was well-separated from most of the background 
population.  BHSM obtained spectroscopy of candidate low-mass members 
to search for signatures of youth and membership, such as Li 
absorption, strong H$\alpha$ emission, and pre-main-sequence spectral features.
Although their observations were sensitive to 
unreddened young brown dwarfs with masses of $\gtrsim0.03$~$M_{\odot}$ (M8-M9),
they found objects down to only $\sim0.08$~$M_{\odot}$ (M6-M6.5). 
However, because the observations were conducted in the optical and the
members become redder with later types, the survey was complete to 
progressively lower levels of extinction with decreasing mass.
Consequently, it was unclear whether the absence of brown dwarfs in the 
BHSM study was statistically significant. 

To measure an IMF for Taurus that is representative at masses of 
$\gtrsim0.02$~$M_{\odot}$, 
I have obtained spectroscopy of low-mass candidates identified by LR98 and 
have searched for brown dwarfs at low masses and moderately high reddenings by 
combining deep $I$ and $z\arcmin$ imaging of the BHSM fields with IR photometry 
from the Two-Micron All-Sky Survey (2MASS). The techniques employed by
Luhman et al.\ (2000) and references therein are used to calculate the IMF 
for this area. The Taurus IMF is found to have the same shape as the 
Trapezium IMF from 0.1-1~$M_{\odot}$, but with moderately fewer stars above 
1~$M_{\odot}$ and a significant deficit of brown dwarfs.  The
implications of these results for theories of the IMF are discussed. 

\section{Observations and Data Analysis}
\label{sec:obs}

\subsection{Spectroscopy}
\label{sec:obs1}

I selected for IR spectroscopy 16 of the 17 low-mass candidates identified 
by LR98 in $JK_{\rm s}$ imaging of a 
$10\arcmin\times10\arcmin$ field towards 
the area of highest extinction in the L1495 dark cloud.
Additional spectra were obtained for the dwarf spectral type standards 
GL~382 (M2V), GL~381 (M2.5V), GL~51 (M5V), LHS~2243 (M8V), LHS~2065 (M9V),
BRI~0021-0214 ($\geq$M9.5), and 2MASP~J0345432$+$254023 (L0V) and the 
optically-classified pre-main-sequence sources V410~X-ray~3 (M6), 5a (M5.5), 
and 6 (M5.5). The observations were performed with the near-IR long-slit 
spectrometer CRSP at the Kitt Peak 4~meter telescope on the nights of 1998 
November 30 and December 1. The 75~l~mm$^{-1}$ grating was used with the 
$1\farcs9$ slit, providing coverage of the entire $K$-band with a two-pixel 
resolution 
of $R=\lambda/\Delta\lambda=300$. Wavelength calibration was performed with
HeNeAr lamp spectra. LR~1 and LR~15/V410~Anon~14 were also observed 
at a spectral resolution of $R=1200$ with the IR 
spectrometer FSpec (Williams et al.\ 1993) at the Multiple Mirror Telescope 
(MMT) on Mount Hopkins on 1997 October 14.
The remaining observing and data reduction procedures for both instruments were 
similar to those described by Luhman et al.\ (1998b).
I also obtained optical spectra of 
LR~10/V410~Anon~3, LR~13/V410~Anon~1, LR~14/V410~Anon~2, LR~15/V410~Anon~14,
V410~X-ray~8b, 8d, and 8e, where the latter three stars are 
candidate counterparts to X-ray sources in the study of L1495 by 
Strom \& Strom (1994).  These measurements were made with the Red Channel 
Spectrograph at the MMT on 1997 November 27-29 and are identical to 
observations of stars in IC~348 by Luhman et al.\ (1998b) during the same 
nights.

\subsection{Photometry}
\label{sec:obs2}

For optical imaging, I selected the regions observed by BHSM towards 
the L1495, B209, L1529, and L1551 dark clouds (see their Figure~1). 
Images of these fields were obtained with the four shooter camera at the
Fred Lawrence Whipple Observatory 1.2~m telescope on 1999 October 12 and 13
under photometric conditions.  The instrument contained four $2048\times2048$
CCDs separated by $\sim45\arcsec$ and arranged in a $2\times2$ grid. After
binning $2\times2$ during readout, the plate scale was $0\farcs67$~pixel$^{-1}$.
Two positions separated by $40\arcsec$ in right ascension and declination 
were observed towards L1495, B209, L1529, and contiguous 
north and south sections of L1551.
At each position, images were obtained at $I$ ($\lambda_{eff}\sim8100$~\AA)
and $z\arcmin$ ($\lambda_{eff}\sim9100$~\AA) with exposure
times of 1 and 20~min for each filter.  The images were bias subtracted,
divided by dome flats, registered, and combined into one image at each
band. Image coordinates and photometry were measured with DAOFIND and PHOT
under the IRAF package APPHOT.
Aperture photometry was extracted with a radius of four pixels. The background 
level was measured in an annulus around each source and subtracted from
the photometry, where the inner radius of the annulus was five pixels and the
width was one pixel.
The photometry was calibrated in the Cousins $I$ system through observations
of standards across a range of colors (Landolt 1992). Because
the $I$ filter for these observations was similar to Cousins $I$, the
color transformation was small. The $z\arcmin$ filter is similar to that
of the Sloan Digital Sky Survey (Fukugita et al.\ 1996). The $z\arcmin$ data
were calibrated by the standard star with the most neutral colors ($V-I=0.2$)
and assuming $I-z\arcmin=0$. This crude calibration is sufficient for
this study since the $z\arcmin$ data are used only in identifying likely
background stars and young low-mass candidates through the relative 
$I-z\arcmin$ colors.  Saturation occurred near $I\sim12.5$. The completeness 
limits were $I\sim21$ and $z\arcmin\sim19.5$ as inferred from a comparison
of the data to the number of stars as a function of magnitude in the Galactic
models of Bahcall \& Soneira (1981). Typical photometric uncertainties 
are 0.04~mag at $I=20$ and $z\arcmin=18.5$ and 0.1~mag at $I=21$ and 
$z\arcmin=19.5$.  
The plate solution was derived from coordinates of sources observed 
in the 2MASS Spring 1999 Release Point Source Catalog that appeared in
the optical images and were not saturated. 

The centers and dimensions of the fields from which photometry is extracted 
are $(\alpha,\delta) (2000)=
(4^{\rm h}18^{\rm m}36\fs65$, $28\arcdeg23\arcmin30\farcs5$) and 
$23\farcm42\times23\farcm42$ for L1495,
$(\alpha,\delta) (2000)=
(4^{\rm h}14^{\rm m}12\fs15$, $28\arcdeg10\arcmin51\farcs5$) and 
$23\farcm42\times23\farcm42$ for B209,
$(\alpha,\delta) (2000)=
(4^{\rm h}32^{\rm m}31\fs1$, $24\arcdeg23\arcmin37\farcs5$) and 
$23\farcm1\times22\farcm5$ for L1529,
and $(\alpha,\delta) (2000)=
(4^{\rm h}31^{\rm m}47\fs8$, $18\arcdeg09\arcmin35\farcs0$) and 
$23\farcm18\times46\farcm83$ for L1551.
2MASS photometry is currently unavailable for the northern 70\% of the
L1551 field (north of $\delta=18\arcdeg00\arcmin06\farcs0$). Out of nearly
2000 2MASS sources within these four fields, 66 objects were not measured
in the optical data because they fell within gaps between the detectors, 
were contaminated by bad pixels, or were not point sources (e.g.,
galaxies).  These 2MASS sources are excluded from the study. In addition,
a known asteroid that was flagged in the 2MASS database was rejected.
In the areas
covered by 2MASS, stars saturated in the optical data were included in the
compilation of sources through their 2MASS photometry and coordinates. But
in the portion of L1551 not available in 2MASS, no photometry is available for
saturated optical sources, and thus they are omitted. The exceptions are
V826~Tau, XZ~Tau, HL~Tau, V827~Tau, V710~Tau~A and B, which are young stars with
previously measured photometry. Two additional young stars that fall in the 
L1551 field and lack 2MASS data are L1551/IRS5 and LkH$\alpha$~358. Because the 
former is broad and nebulous in the optical data, photometry and coordinates 
were not measured for it in this work. For these eight stars, IR photometry 
is taken from Kenyon \& Hartmann (1995).
The recently discovered low-mass objects MHO~4, MHO~5, MHO~6, MHO~7, and 
MHO~9 (BHSM) lack 2MASS photometry or other previous IR data.
Photometry and coordinates for spectroscopically confirmed young members
within the imaged fields of Taurus are listed in Table~1.
When available, the coordinates in Table~1 are those measured from the
optical data. For the other stars -- 2MASS sources that were saturated or 
too red to be detected in the optical images -- the 2MASS coordinates are given.

\section{Individual Source Characteristics}

\subsection{Spectral Types}
\label{sec:sptypes}

The targets in the spectroscopic sample can be foreground stars, background
stars, or young low-mass members of Taurus. Foreground M dwarfs are 
identified by the lack of reddening in their spectra and colors and by the 
strong absorption in the optical Na and K transitions (Luhman et al.\ 1998a, 
1998b).  The star LR~10/V410~Anon~3 exhibits these properties with a spectral 
type of M5V, which is consistent with the classification as a foreground
M4.5V star by Strom \& Strom (1994).
Other previously known foreground M stars in L1495 include V410~Anon~12 
and 17 (Strom \& Strom 1994; BHSM). Note that V410~Anon~3 was mistakenly 
omitted as an alternate designation for LR~10 in Table~2 of LR98. In addition, 
LR~11 and LR~17/V410~Anon~27 are in fact the same star. As discussed by Strom 
\& Strom (1994), V410~Anon~9 may be a field star close to and behind the L1495
cloud. It is taken as a background star in the remaining analysis. The question
of membership of this star has no bearing on the analysis of the IMF because 
it has a reddening higher than the limit used in defining the IMF sample. 

Young late-type members of Taurus are expected to have
strong absorption in TiO, VO, and H$_2$O at optical and near-IR wavelengths.
Thus, they should be easily identified through low-resolution spectroscopy from
$I$ through $K$.  On the other hand, background stars are predominantly
giants and early-type stars. These stars exhibit featureless spectra 
at low resolution in the optical and IR, except for giants which have strong
CO absorption at 2.3~$\mu$m. The following targets from the 
spectroscopic sample are classified as background stars:
V410x8b (RC, C), 8d (RC), 8e (RC, LR98), LR~2 (C), LR~3/V410~Anon~19 (C), 
LR~4/V410~Anon~21 (C), LR~5-8 (C), LR~9/V410~Anon~26 (C), LR~12 (C), 
LR~13/V410~Anon~1 (RC), LR~14/V410~Anon~2 (RC), LR~15/V410~Anon14 (BHSM, RC, F),
and LR~17/V410~Anon~27 (C).
The classifications are from the new data with Red Channel (RC), CRSP (C), 
and FSpec (F) and from the previous work of BHSM and LR98.

A spectral type of K4-K5 is measured for LR~1 from the FSpec data in the manner
described by LR98. This star cannot be a foreground star because of its 
reddened colors. It is too bright for its spectral type to be a 
background star (i.e., above the main sequence), therefore it is taken as a
young member of Taurus. Spectral types for LR~1 and all other known young 
sources within the four Taurus fields are listed in Table~1. 
 
\subsection{Extinctions}
\label{sec:colors}

Extinctions for the known young stars in the four Taurus fields are
now estimated. 

In the following analysis,
standard dwarf colors are taken from the compilation of Kenyon \& Hartmann 
(1995) for types earlier than M0 and from the young disk populations 
described by Leggett (1992) for types of M0 and later. The IR colors from
Leggett (1992) are on the CIT photometric system. Most of the IR data in 
Table~1 are from the 2MASS survey, where the photometric system is similar to 
CIT. The IR colors of the standards and the eight young stars taken from 
Kenyon \& Hartmann (1995) are transformed from Johnson-Glass to CIT 
(Bessell \& Brett 1988) during the analysis, although
the colors of those young stars remain in Johnson-Glass in Table~1. 
Reddenings are calculated with the extinction law of Rieke \& Lebofsky (1985).

In the common method for estimating the reddening towards a young star, 
a color excess is calculated by assuming the intrinsic color, typically 
that of a standard dwarf at the spectral type in question. To ensure that
the color excess reflects only the effect of reddening, 
contamination from short or long wavelength excess emission is minimized
by selecting colors between the $R$ and $H$ bands. Because $R-I$ is 
less susceptible to excess emission than $J-H$, the former is used 
for measuring reddening in this study when it is available. The $R-I$ colors 
of 35 stars in Table~1 are provided by BHSM, Strom \& Strom (1994), and
Kenyon \& Hartmann (1995). The reddenings computed by BHSM for their 
late M objects are slightly lower than the values reported here because
of the differing references for standard dwarf $R-I$ colors between the two 
studies.  Although dwarf colors have been assumed here, the intrinsic 
$R-I$ colors of young stars do appear to depart from dwarf
values for M4-M5 types (Luhman 1999).

Meyer, Calvet, \& Hillenbrand (1997) estimated the locus of intrinsic $J-H$ 
and $H-K$ colors for a sample of classical T~Tauri stars (CTTS) by 
dereddening the observed colors with the extinctions measured from $R-I$.
The locus had an origin that fell near the intrinsic colors
of an M0 dwarf and it extended to redder $J-H$ and $H-K$ with a well-defined 
slope, a behavior that was reproduced by models of star-disk systems.
When the same experiment is repeated for the sources with $R-I$ in this study,
the stars are distributed around the CTTS locus of Meyer et al.\ (1997) in 
a similar fashion as in Figure~4 of LR98 for L1495. However, the scatter 
is larger than that found by Meyer et al.\ (1997), probably because the sample 
here is not as homogeneous in spectral type or as well-studied photometrically. 

To calculate reddenings for each of the 14 stars in Table~1 that has a 
spectral type but lacks an $R-I$ measurement, the $J-H$ and $H-K$ colors are 
dereddened to a CTTS locus where the 
origin coincides with the dwarf colors for the star's spectral type. The locus 
is given the slope measured by Meyer et al.\ (1997) for M0 stars, which they 
predicted to remain relatively constant with spectral type.  This method 
assumes that the central stars of young mid- and late-M stars have dwarf-like 
$J-H$ and $H-K$ colors, which is suggested by the work of Luhman (1999). 
Most of the stars dereddened in this fashion 
have extinctions that are nearly the same as those derived by
simply assuming dwarf colors without including the CTTS locus, i.e., 
they have little IR excess emission.
An exception is PSC04154+2823, which exhibits an excess in $H-K$ that is 
larger than expected from a reddened CTTS. To estimate the extinction for
this object, the $J-H$ color is dereddened to a value of 1.1, which is 
roughly the maximum intrinsic color of the CTTS locus. 
In the compilation of Kenyon \& Hartman (1995), the binary V710~Tau 
($3\farcs24$, Leinert et al.\ 1993) is resolved in the IR photometry 
but not in the optical data. Because the intrinsic optical colors of the
components probably differ significantly given their spectral types, the 
IR colors are used in measuring the extinctions towards each component.
Reddenings and luminosities are not calculated for the three sources lacking 
spectral types and the class~I object L1551/IRS5.

\subsection{Effective Temperatures and Bolometric Luminosities}
\label{sec:hr}

For reasons described in previous studies (e.g., Luhman 1999), $I$ and $J$
are the preferred bands for measuring bolometric luminosities of young 
low-mass stars. When $J$ is available, the luminosities in Table~1 are computed 
from standard dwarf bolometric corrections (see Luhman 1999), the dereddened 
$J$-band measurements, and a distance modulus of 5.76 (Wichmann et al.\ 1998).
For the five members that lack IR data, luminosities are estimated from
the $I$-band photometry obtained in this study. 
Because HL~Tau, PSC04154+2823, and MHO~1 have strong excess emission in the
near-IR, the luminosities inferred from $J$ have large uncertainties.
In addition, such sources are often highly variable, as found for PSC04154+2823 
when the photometry of this work and BHSM are compared ($\Delta I=1.66$). 

Spectral types of M0 and earlier are converted to effective temperatures with 
the dwarf temperature scale of Schmidt-Kaler (1982).  For spectral types later 
than M0, the adopted temperature scale is that developed by Luhman (1999)
for use with the evolutionary models of Baraffe et al.\ (1998).  Detailed 
discussions of the temperature scales and evolutionary models for young
low-mass stars are found in LR98, Luhman (1999), and Luhman et al.\ (2000). 

\section{The Taurus Stellar Population}

\subsection{Membership and Completeness}
\label{sec:member}

\subsubsection{Separation of Field Stars and Candidate Low-Mass Members}
\label{sec:sep}

To distinguish cool, low-mass cluster members from background stars in a 
color-magnitude diagram, the photometric bands should be selected such that the
directions of decreasing mass and increasing extinction differ 
as much as possible. For example, in a plot of $J-H$ versus $H$, 
members of a population will move vertically down the diagram with lower 
mass and cooler temperatures because the near-IR colors of most spectral 
types fall within a small range of values.  Meanwhile, extinction moves
stars down and to the right at a relatively shallow angle in this diagram.
As a result, cluster members and reddened background stars are mixed together
and cannot be separated with IR colors alone. However, in a color-magnitude
diagram that includes at least one optical band (e.g., $R-I$, 
$I-z\arcmin$, $I-J$), the color increases rapidly with later spectral types
and the reddening vector has a steep slope downward. In such a diagram, 
the intersection of reddened background stars and low-mass members is minimized,
allowing for the efficient rejection of most of the field stars. 

\begin{figure}
\plotone{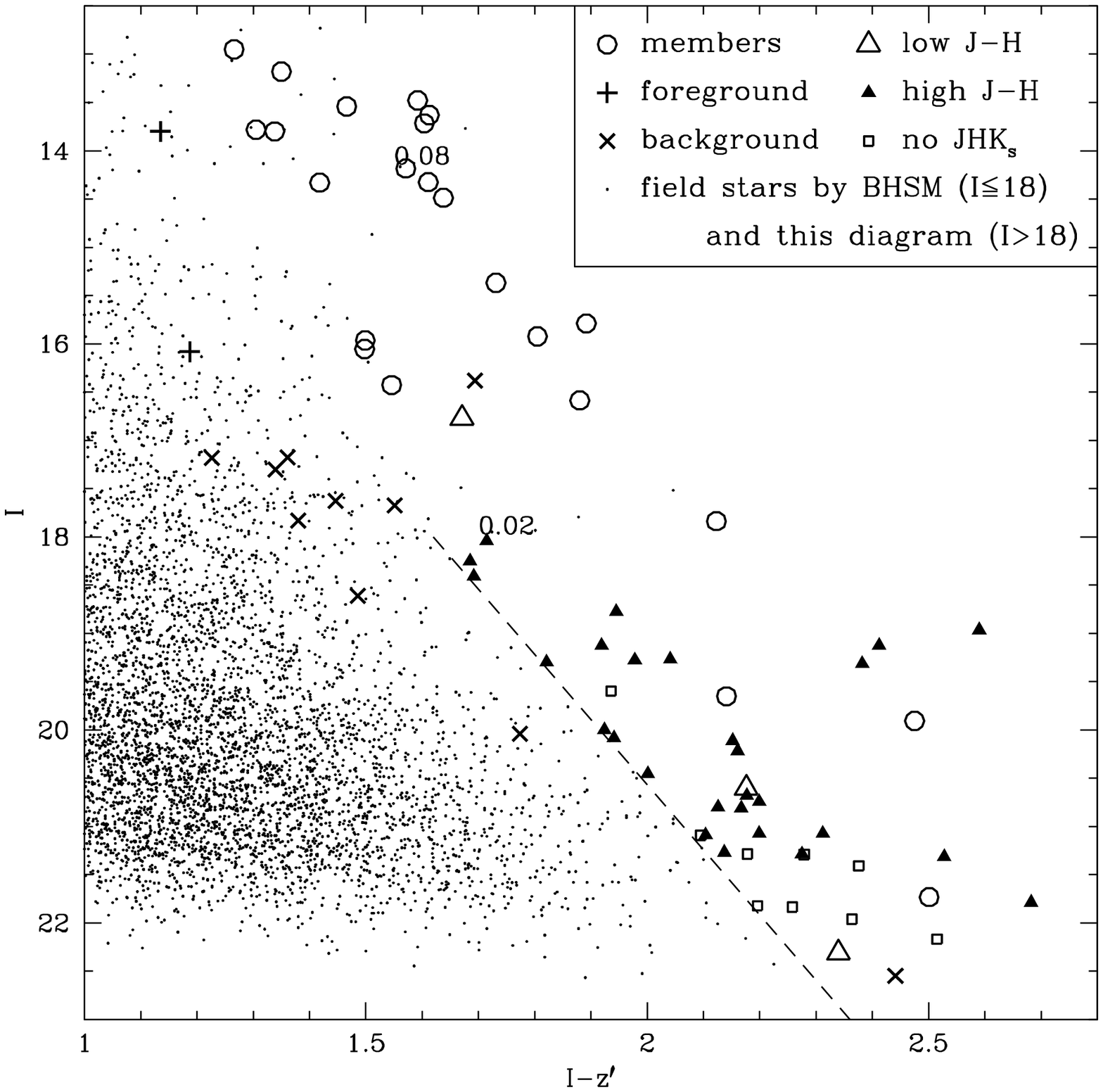}
\caption{
$I-z\arcmin$ vs.\ $I$ for fields surrounding the L1495, B209, L1529, 
and L1551 dark clouds in the Taurus star forming region.  
Stars that have been spectroscopically identified as Taurus members, 
foreground stars, and background stars are indicated.
At $I\leq18$, all other stars except for one object (open triangle) are
field stars by the spectroscopy and $R-I$ vs.\ $I$ diagram of BHSM. 
At $I>18$, Taurus members with masses of $\geq0.02$~$M_{\odot}$ and ages of 
$\leq10$~Myr are expected to fall above the dashed reddening vector. Any
stars below this line are labeled as field stars. 
The stars above the reddening vector are categorized as having low $J-H$,
high $J-H$, or no $JHK_s$ photometry. 
Objects with low $J-H$ are within the reddening limit of $A_H\leq1.4$ that 
defines the sample from which the IMF is calculated. Sources without $JHK_s$
photometry are within the portion of the L1551 not currently available in 2MASS.
The unreddened positions of 0.08 ($\sim$M6.5) and 0.02~$M_{\odot}$ ($\sim$M9)
are shown for an age of 1~Myr (Baraffe et al.\ 1998).
}
\label{fig:iz}
\end{figure}

\begin{figure}
\plotone{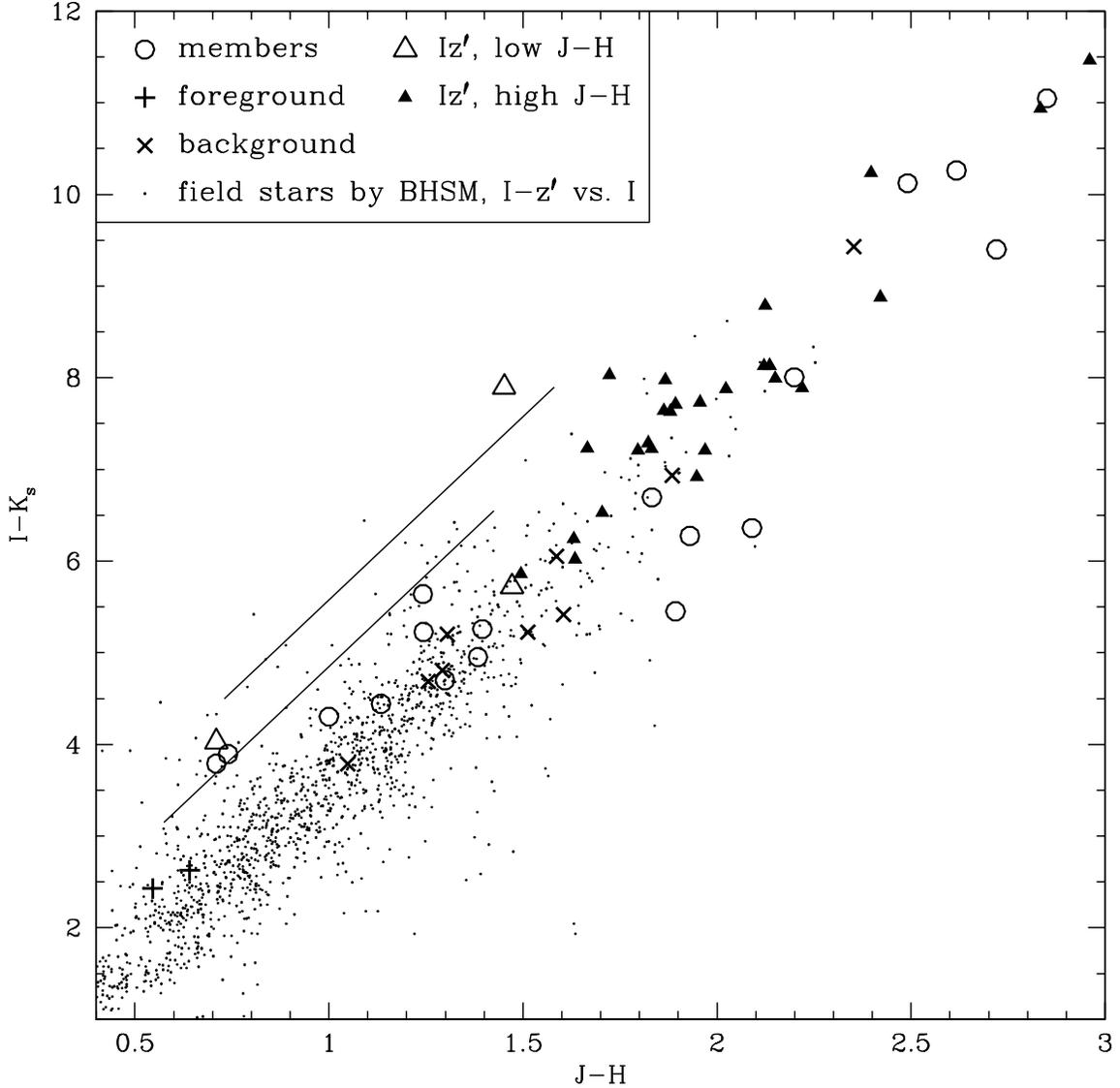}
\caption{
$J-H$ vs.\ $I-K_s$ for fields surrounding the L1495, B209, L1529,
and L1551 dark clouds in the Taurus star forming region.  
Stars that have been spectroscopically identified as Taurus members, 
foreground stars, and background stars are indicated.
Objects that are field stars by the $I$ and $z\arcmin$ data in 
Figure~\ref{fig:iz} and by the spectroscopy and $RI$ data of BHSM are
also shown.
The remaining stars that are not rejected as field stars through those
means are categorized as having low $J-H$ or high $J-H$. 
Objects with low $J-H$ are within the reddening limit of $A_H\leq1.4$ that 
defines the sample from which the IMF is calculated.
The lower and upper lines represent the reddening vectors from $A_H=0$-1.4
for spectral types of M6.5 and M9 (Leggett 1992), which correspond to 0.08 and
0.02~$M_{\odot}$ for an age of 1~Myr (Baraffe et al.\ 1998).
}
\label{fig:jhik}
\end{figure}

For the above reasons, BHSM used an optical color-magnitude diagram to
search for new low-mass members of Taurus in fields towards L1495, B209, L1529,
and L1551. 
At stellar masses, spectroscopy was obtained for nearly all candidate members;
only a few objects near the locus of Taurus members were not observed 
spectroscopically. 
BHSM obtained spectra for all but one of the brown dwarf candidates
at $I\leq18$ in the diagram of $R-I$ versus $I$ (their Figure~2). 
The object lacking a spectrum is a likely brown dwarf 
(see \S~\ref{sec:candidates}) and is shown as an open triangle in the 
various color-color and color-magnitude diagrams.
Thus, in the $I-z\arcmin$ versus $I$ and $J-H$ versus $I-K_s$ diagrams in 
Figs.~\ref{fig:iz} and \ref{fig:jhik}, after plotting this brown dwarf
candidate and the spectroscopically confirmed members, all other stars 
at $I\leq18$ can be labeled as field stars.  Because the spectroscopy
is not 100\% complete for the candidates at stellar masses, a few of 
the objects taken as field stars may be young stars.

I can now use the color-color and color-magnitude diagrams presented here 
to separate background stars and candidate 
low-mass members that are beyond the sensitivity of the BHSM study ($I=18$-21).
To determine the regions in Figs.~\ref{fig:iz} and \ref{fig:jhik} where
low-mass Taurus members should appear, I must plot the intrinsic 
colors and magnitudes expected for young brown dwarfs at masses of 0.02 
and 0.08~$M_{\odot}$.  
The $I-z\arcmin$ colors for these masses are estimated in the following manner. 
The combination of the models of Baraffe et al.\ (1998) and the compatible 
temperature scale of Luhman (1999) indicate that brown dwarfs at masses of 
0.02 and 0.08~$M_{\odot}$ and an age of 1~Myr should have spectral types near
M9 and M6.5, respectively. From the locus of reddened background stars in 
Figure~\ref{fig:jhik}, I measure a reddening slope of $E(I-K_s)/E(J-H)=4.0$.  
By combining this slope with the dwarf colors of Leggett (1992) for 
spectral types of M9 and M6.5, I plot the reddening vectors for 0.02 and 
0.08~$M_{\odot}$ from $A_H=0$-1.4 in Figure~\ref{fig:jhik}.
The $E(I-K_s)/E(J-H)$ slope is also combined with $E(J-H)=0.107$~$A_V$ (Rieke 
\& Lebofsky 1985) and $A_{K_s}=0.116$~$A_V$ (extrapolated from $A_K$ of 
Rieke \& Lebofsky 1985) to derive $A_I=0.544$~$A_V$. After measuring a slope of 
$E(I-z\arcmin)/E(J-H)=0.75$ from the distribution of field stars in $I-z\arcmin$
versus $J-H$, I arrive at $E(I-z\arcmin)=0.08$~$A_V$.
From this extinction relation and the reddenings in Table~1, the intrinsic 
$I-z\arcmin$ colors of the M6-M6.25 Taurus members are calculated. 
These dereddened $I-z\arcmin$ colors are then extrapolated to M6.5 and M9 
($I-z\arcmin=1.6$ and 1.75) by using the variation in $I-Z$ from M6 to 
M9 for objects in the field and the Pleiades open cluster 
(Steele \& Howells 2000; Zapatero Osorio et al.\ 1999).
The magnitudes at $I$ for 0.02 and 0.08~$M_{\odot}$ at an age of 1~Myr 
are calculated by combining the bolometric corrections and distance modulus
from \S~\ref{sec:hr} with the luminosities predicted by Baraffe et al.\ (1998).
The resulting positions for these two masses are shown in Figure~\ref{fig:iz}.
Most stars in these Taurus fields have ages of $\lesssim3$~Myr, with a few as 
old as 10~Myr (\S~\ref{sec:tau}). Therefore, to obtain a census that is
complete down to 0.02~$M_{\odot}$, the boundary in Figure~\ref{fig:iz}
used in selecting low-mass candidates must include objects at 0.02~$M_{\odot}$
with ages as old as 10~Myr.  From 1 to 10~Myr, the evolutionary models of 
Baraffe et al.\ (1998) and Burrows et al.\ (1997) predict a decrease in 
luminosity of 0.5 and 0.9~mag, respectively.  Thus, to be conservative in
the rejection of field stars, a reddening vector 
is placed at 1~mag below the position of 0.02~$M_{\odot}$ in 
Figure~\ref{fig:iz}; any low-mass members with masses above 0.02~$M_{\odot}$ 
should fall above this line.  The nature of stars above the reddening vector 
will be examined in more detail in \S~\ref{sec:candidates}. 
Stars below this vector are rejected as field stars for the purposes of 
calculating the IMF down to 0.02~$M_{\odot}$. However, substellar
members of Taurus with masses of $\lesssim0.01$~$M_{\odot}$ ($\gtrsim$L0) 
could fall below the reddening vector because the $I-z\arcmin$ color 
saturates at early L types (Steele \& Howells 2000).
As demonstrated in Figure~\ref{fig:iz}, most of the targets identified as
foreground and background stars through spectroscopy could have been rejected 
by deep $I$ and $z\arcmin$ photometry alone had it been available.

\subsubsection{Selection of the Reddening Limit of the IMF Sample}
\label{sec:select}

Because the mass and reddening vectors are roughly perpendicular in a
near-IR color-magnitude diagram and because all of the late-type members 
should have similar intrinsic IR colors, completeness 
in mass and reddening are readily evaluated with near-IR data. Thus, 
by combining the 2MASS data with the membership information from the 
spectroscopy and the optical color-magnitude diagrams of BHSM and this work, 
I can determine the optimum reddening limit for the sample used in calculating 
the IMF. This reddening limit will be high enough to include a large 
number of cluster members while low enough to achieve completeness to very 
low masses.  In Figure~\ref{fig:hjh}, the near-IR color-magnitude diagram is 
generated from the 2MASS data for the four Taurus fields, with the exception 
of the portion of L1551 that was unavailable in 2MASS (see \S~\ref{sec:obs2}). 
The membership status is established for most of the faint 
sources at $J-H\leq1.5$. Because this color corresponds to a reddening
of $A_H=1.4$ for young brown dwarfs ($J-H\sim0.65$; Luhman 1999), 
I select a reddening limit of $A_H\leq1.4$ for the sample from which the 
IMF is computed. If a young brown dwarf 
with $A_H\leq1.4$ exhibited an excess at $J-H$ arising from emitting
circumstellar material, it could fall in the region of high $J-H$
in Figure~\ref{fig:hjh} where membership status is uncertain for most objects. 
However, the fraction of stellar members that have such large excesses at
$J-H$ is very small. Furthermore, the young brown dwarfs studied by Luhman 
(1999) in IC~348 appear to have negligible excess emission in $J-H$.

\subsubsection{Any Brown Dwarfs within the Reddening Limit?}
\label{sec:candidates}

\begin{figure}
\plotone{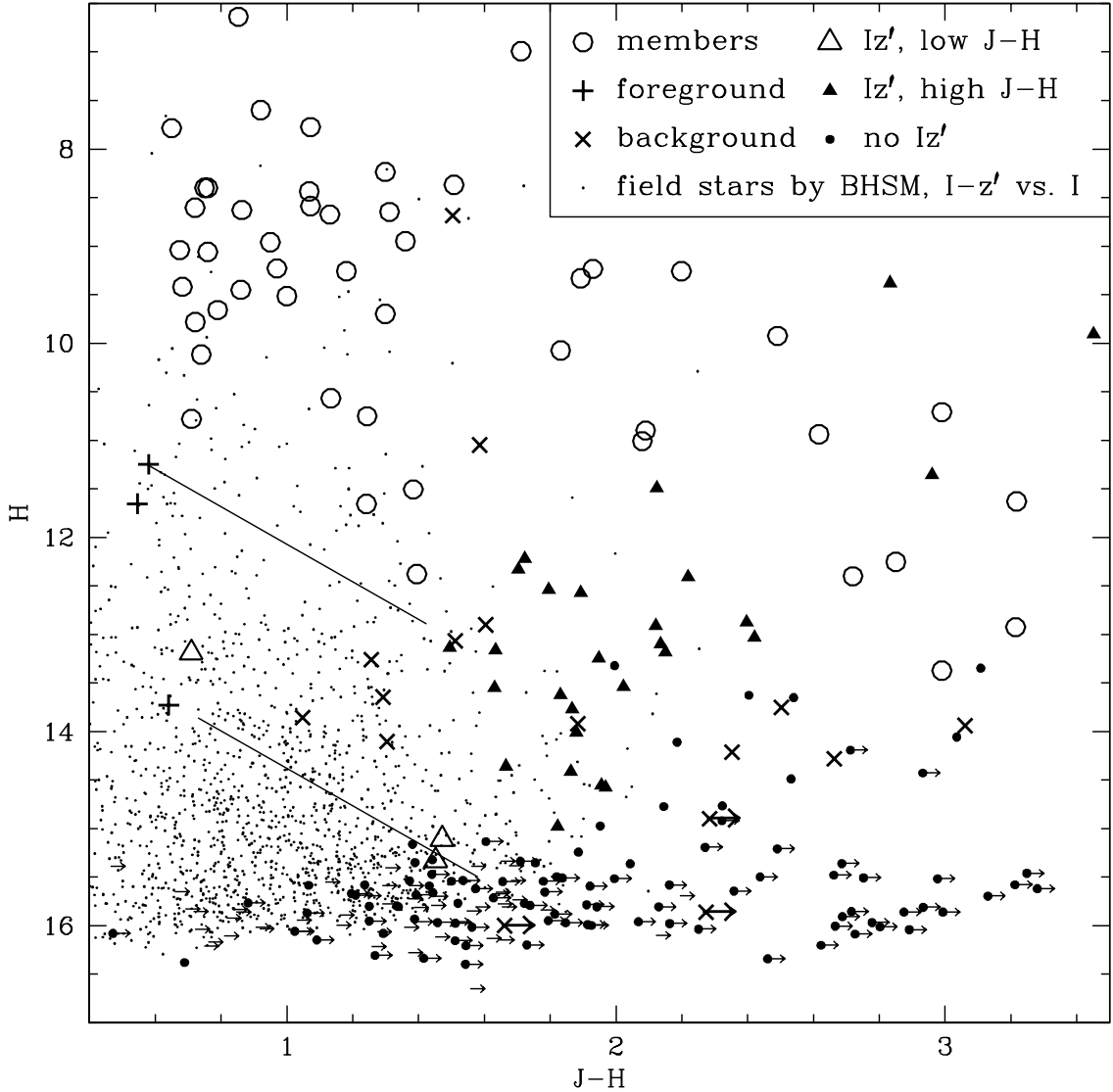}
\caption{
$J-H$ vs.\ $H$ from the 2MASS survey for fields surrounding the
L1495, B209, L1529, and L1551 dark clouds in the Taurus star forming region,
with the exception of portion of L1551 not currently available in 2MASS. 
Stars that have been spectroscopically identified as Taurus members, 
foreground stars, and background stars are indicated.
Objects that are field stars by the $I$ and $z\arcmin$ data in 
Figure~\ref{fig:iz} and by the spectroscopy and $RI$ data of BHSM are
also shown.
The remaining stars that are not rejected as field stars through those means 
are categorized as having low $J-H$, high $J-H$, or no $I$ and $z\arcmin$ data. 
Objects with low $J-H$ are within the reddening limit of $A_H\leq1.4$ that 
defines the sample from which the IMF is calculated.
The upper and lower lines represent the reddening vectors from $A_H=0$-1.4 
for 0.08 ($\sim$M6.5) and 0.02~$M_{\odot}$ ($\sim$M9) for an age of 1~Myr 
(Baraffe et al.\ 1998).
}
\label{fig:hjh}
\end{figure}

I now determine whether there are any new low-mass candidates within the
reddening limit that is used for the IMF measurement. In Figs.~1-4, 
stars that have not been rejected as field stars in \S~\ref{sec:sep}
and are not known members
are divided into three categories: 
stars detected only in the 2MASS data (no $I$ and $z\arcmin$), 
optical sources with high reddening (high $J-H$), 
and optical sources with low reddening (low $J-H$). For the former sources, 
the membership status cannot be unambiguously determined because only IR
data is available. However, this is not
important for the IMF calculation because these objects are either 
fainter than the $H$ magnitude of the desired completeness limit of 
0.02~$M_{\odot}$ or are redder than the extinction limit of $A_H\leq1.4$,
as shown in Figure~\ref{fig:hjh} and discussed further in \S~\ref{sec:complete}.
The high reddening category consists of sources that fall to the right of
the $A_H\leq1.4$ reddening vectors in Figure~\ref{fig:hjh}; these stars
are beyond the extinction limit for the IMF and therefore are not considered
further. Finally, the low reddening objects are defined as those
between the $A_H\leq1.4$ reddening vectors for 0.02 
and 0.08~$M_{\odot}$ or below the vector for 0.02~$M_{\odot}$ 
in Figure~\ref{fig:hjh}. Because the three objects in this category are within
the reddening limit of the IMF sample, their membership status must be
examined to determine whether they should be added to the IMF. The brightest
of these three sources is 2MASSs~J0418511+281433, 
the brown dwarf candidate of BHSM mentioned in \S~\ref{sec:sep}.
The 2MASS and optical photometry for this object are 
$H=13.19$, $J-H=0.71$, $H-K_s=0.44$, $I=16.77$, and $I-z\arcmin=1.67$.
As shown in Figure~2 of BHSM and in Figs.~\ref{fig:iz} and 
\ref{fig:jhik}, 2MASSs~J0418511+281433 has the colors and magnitudes 
expected for a young brown dwarf. 
The second brightest source is 2MASSs~J0432086+242213, which has 
$H=15.11$, $J-H=1.47$, $H-K_s=0.22$, $I=20.61$, and $I-z\arcmin=2.18$.
This object does not have a high enough ratio of $I-K_s$ to $J-H$ to be a 
late-type member of Taurus, as demonstrated in Figure~\ref{fig:jhik}.
The third object, 2MASSs~J0417504+281440, has 
$H=15.33$, $J-H=1.45$, $H-K_s=0.92$, $I=22.31$, and $I-z\arcmin=2.34$.
This source has $J-H$ and $I-K_s$ colors 
that are consistent with a late M type (reddest of the three open triangles
in Figure~\ref{fig:jhik}). However, because 
it falls at the detection limit in both the optical and IR data, its
positions in the color-color and color-magnitude diagrams are uncertain.
This source is taken to be a field
star for the remainder of this work; the membership status of 
this one object is not important for the conclusions of this study. 
In summary, there is one compelling brown dwarf candidate lacking spectroscopy
and within the reddening limit of $A_H\leq1.4$ used for the IMF in L1495, 
B209, L1529, and the portion of L1551 currently available through 2MASS.

\subsubsection{Special Considerations for L1551}

For the portion of the L1551 field that is not included in the 2MASS photometry,
the identification of low-mass candidates and estimation of completeness must 
be performed with the optical data alone.  The objects in this region that 
are not rejected as field stars by the $I-z\arcmin$ versus $I$ diagram 
are labeled with open boxes in Figure~\ref{fig:iz}.
Without IR data, the completeness is not assured for high reddenings at
substellar masses. Therefore, rather than use a reddening limit of 
$A_H\leq1.4$ to calculate the IMF as done in the other fields, a smaller
range of extinctions must be considered for this part of L1551. 
Because 12 of the 14 known young stars in L1551 have $A_H=0$-0.5, 
$A_H\leq0.5$ is taken as the reddening limit for the calculation of the
IMF in this part of L1551. For the IMF to be representative for masses of
$\leq0.02$~$M_{\odot}$, low-mass candidates must be identified down to 
the magnitude of a 0.02~$M_{\odot}$ brown dwarf at an age of 10~Myr with a
reddening of $A_H=0.5$, which corresponds to $I=20.6$. This magnitude was
computed by combining the luminosities and temperatures predicted by
Baraffe et al.\ (1998) with the temperature scale and bolometric corrections
discussed in \S~\ref{sec:hr}.
There is only one L1551 object above this brightness that is not rejected as 
a field star.  In the other fields covered by 2MASS, all but one
of the stars falling above the reddening vector in Figure~\ref{fig:iz}
were found to be either highly reddened or a field star when the 
optical and IR data were combined. Thus, because it is likely that this 
one object in L1551 would also be rejected in the same manner if IR 
photometry were available, it is not considered a likely brown dwarf.

\subsubsection{The Mass Completeness of the Reddening-Limited Sample}
\label{sec:complete}

\begin{figure}
\plotone{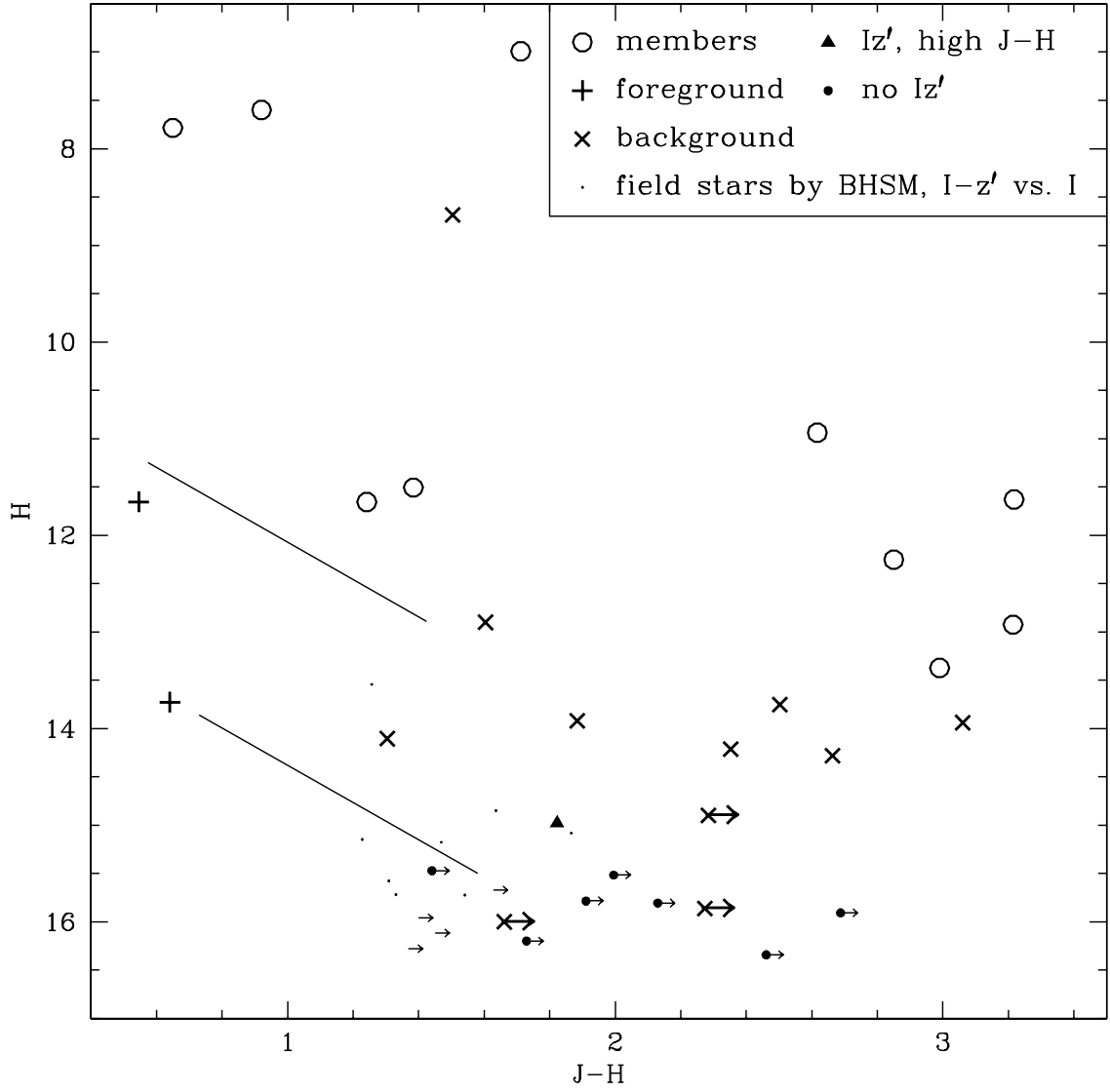}
\caption{
Same as Figure~\ref{fig:hjh}, but restricted to the $10\arcmin\times10\arcmin$
region of the L1495 dark cloud imaged by LR98. 
}
\label{fig:hjhred}
\end{figure}

The reddening-limited sample described here would not be an accurate
reflection of the Taurus population if the average extinctions were a function 
of mass, e.g., young brown dwarfs were more highly reddened than stars. 
In previous magnitude-limited searches for low-mass stars and brown dwarfs 
in star forming clusters (e.g., Luhman 1999; Wilking, Greene, \& Meyer 1999)
no difference in reddening characteristics has been found between the stellar 
and substellar members.  The data in this study can be used as an 
additional test of whether brown dwarfs are hidden by higher amounts 
of extinction than the known stellar population.  Most of the targets 
of the IR spectroscopy were selected from the imaging of LR98 towards a
$10\arcmin\times10\arcmin$ region of the L1495 dark cloud. 
As shown in the IR color-magnitude diagram for this area in 
Figure~\ref{fig:hjhred}, no low-mass members are found at a dereddened
magnitude of $H>10.5$-13 in a sample where the membership status 
is complete to large reddenings ($A_V\lesssim20$). Although this test
suffers from poor number statistics, it is consistent with the result
from studies of other young regions that a reddening-limited measurement 
produces a representative IMF down to substellar masses.

\begin{figure}
\epsscale{0.85}
\plotone{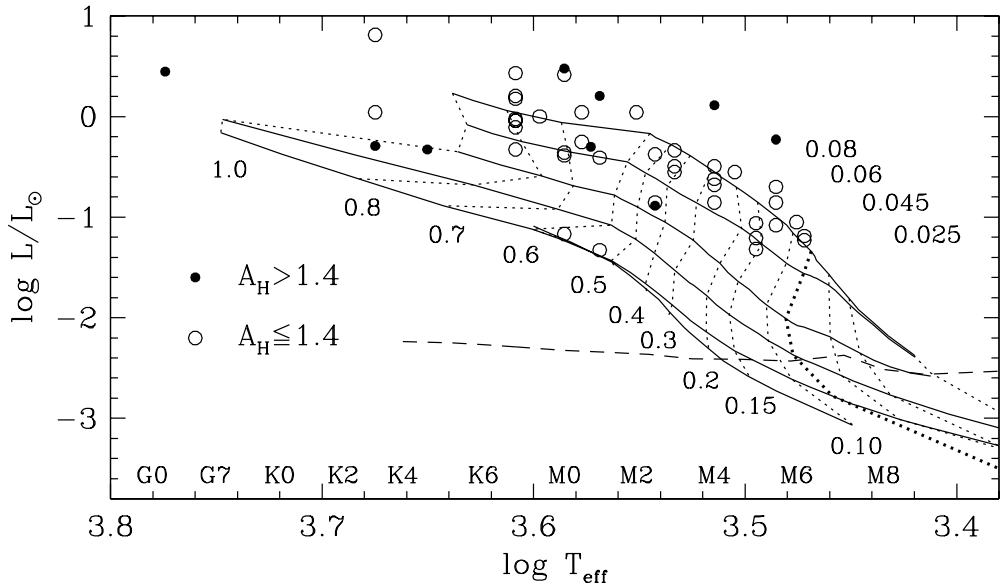}
\caption{
H-R diagram for fields surrounding the
L1495, B209, L1529, and L1551 dark clouds in the Taurus star forming region.
The theoretical evolutionary models of Baraffe et al.\ (1998) are shown, where 
the horizontal solid lines are isochrones representing ages of 1, 3, 10, 30, 
and 100~Myr and the main sequence, from top to bottom.
The dashed line in the H-R diagram represents a dereddened magnitude of $H=14$.
The M spectral types have been converted to effective temperatures with a 
scale that is compatible with these evolutionary models (Luhman 1999), 
which is intermediate between the scales for M dwarfs and giants.
}
\label{fig:hr}
\end{figure}

The sample selected for measuring the IMF is defined by $A_H\leq0.5$ for 
a portion of L1551 and $A_H\leq1.4$ elsewhere. As discussed in the 
previous two sections, out of the objects with optical photometry, there
is one likely brown dwarf within these reddening limits and lacking
spectroscopy; it will be added to the IMF in \S~\ref{sec:tau}. For
stars with only IR photometry, membership cannot be determined. However,
none of these stars are within the reddening ($A_H\leq1.4$) and mass 
($\leq0.02$~$M_{\odot}$) limits for ages of $\leq1$~Myr, as shown in 
Figure~\ref{fig:hjh}. Because the 1 and 3~Myr isochrones are nearly coincident
at 0.02~$M_{\odot}$, this statement holds for an age limit of 3~Myr as well. 
The sample is incomplete only for objects that are simultaneously old (10~Myr) 
and at the reddening and mass limits. Because such objects
occupy a tiny fraction of the mass-reddening-age phase space and 
because most members of these Taurus fields have ages of $\leq3$~Myr (see 
Figure~\ref{fig:hr}), this incompleteness is negligible.
Because there remain $\sim5$ candidate members at 
stellar masses that lack spectra in the color-magnitude diagram of BHSM,
the completeness for brown dwarfs above 0.02~$M_{\odot}$ is as good or better
than that for stars.
Thus, although the census of members may not be 100\% complete in any mass 
interval, it is not biased against brown dwarfs and is representative down to 
a mass of 0.02~$M_{\odot}$.

\subsection{The Initial Mass Function}
\label{sec:imf}

\subsubsection{Taurus}
\label{sec:tau}

\begin{figure}
\epsscale{1}
\plotone{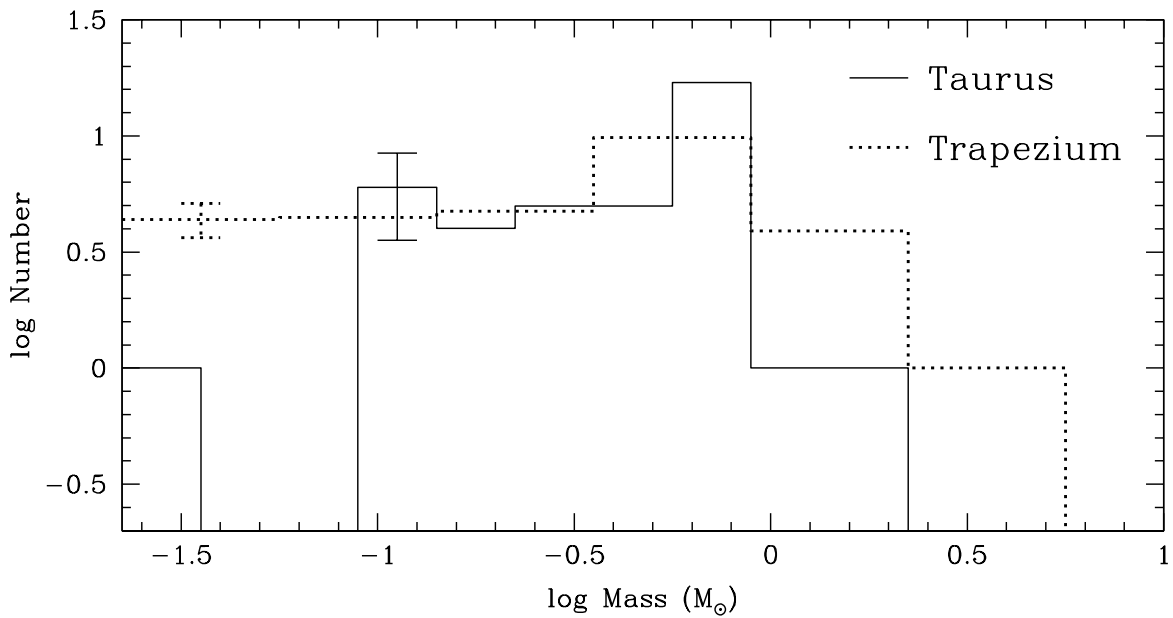}
\caption{
IMFs for reddening-limited samples in the Taurus star forming region and the 
Trapezium Cluster (Luhman et al.\ 2000) as inferred from the evolutionary 
models of Baraffe et al.\ (1998). The IMF for the Trapezium has been 
renormalized for comparison to the Taurus data. 
Both IMFs are representative over the entire mass range shown.
The mass bins for the Trapezium IMF are twice the size of the ones for the
Taurus data (\S~\ref{sec:tau}).
}
\label{fig:imf}
\end{figure}

In the Hertzsprung-Russell (H-R) diagram in Figure~\ref{fig:hr},
the models of Baraffe et al.\ (1998) are plotted with
the Taurus members from Table~1, with the exception of L1551/IRS5, the three 
objects with unknown spectral types, and the two earliest stars. 
These models are used to infer masses for individual sources from their 
estimated temperatures and luminosities. 
For stars in the reddening-limited sample that are above the solar mass track of
Baraffe et al.\ (1998), the prescription of Luhman et al.\ (2000) is followed;
the two stars are placed in one mass bin from log~$M=-0.05$ to 0.35 
(0.89-2.2~$M_{\odot}$). 2MASSs~J0418511+281433 is the one likely Taurus 
member that lacks a measured spectral type and is within the reddening limit 
of the IMF sample. If an age of 1~Myr is assumed for this object, the optical
and IR photometry (\S~\ref{sec:candidates}) imply a mass of 
$\sim0.03$~$M_{\odot}$ with the models of Baraffe et al.\ (1998).
Because the masses of many of the low-mass sources 
in the Trapezium IMF of Luhman et al.\ (2000) were derived through photometry
alone, they used the bin size of $\Delta {\rm log}$~$M=0.4$ at all masses.
For all but one of the sources in the Taurus IMF, on the other hand, more 
precise masses are inferred through 
spectroscopy and placement on the H-R diagram, thus, smaller mass bins of
$\Delta {\rm log}$~$M=0.2$ are used below log~$M=-0.05$.  
The IMF from the reddening-limited sample for the four Taurus fields 
contains 40 sources and is presented in Figure~\ref{fig:imf}.

\subsubsection{Comparison of Taurus and Clusters}

Luhman et al.\ (2000) found that the stellar IMFs in the young clusters
IC~348, $\rho$~Oph, and the Trapezium are consistent with the same mass 
function, while modest variations in the substellar IMFs are possible.
Because of the superior number statistics in the Trapezium sample, 
that IMF is taken to represent star forming clusters in the following 
comparison between Taurus and clusters.

For a meaningful comparison of the IMFs in Taurus and the Trapezium,
it is important that the IMFs have been constructed from samples that are
physically equivalent and relatively primordial.
Thus, the minimum separations of objects in the two samples should 
correspond to the same physical scale and should be less then the 
distance at which disruption through dynamical interactions is significant.
These criteria have been satisfied fairly well;
the IMFs for the Trapezium and Taurus consist of sources with separations 
greater than $0\farcs7$ and $2\arcsec$, respectively, which are equal physical 
distances. Furthermore, these separations are less than the length scales
in the Trapezium ($\sim1\arcsec$) and Taurus ($\sim90\arcsec$) that
mark the transitions from the binary regime to large-scale clustering 
(Gomez et al.\ 1993; Larson 1995; Simon 1997; Nakajima et al.\ 1998; Bate,
Clarke, \& McCaughrean 1998), which in turn should be less than the maximum 
binary separations (Bate et al.\ 1998).

Because the samples in Taurus and the Trapezium have been designed to be
comparable in their binarity and because the same techniques were employed in 
estimating the masses 
for each population, the IMFs for the two regions can be reliably compared. 
In addition, the optimum reddening limit for each sample is $A_H=1.4$ 
(except for L1551), making the two IMFs particularly suitable for comparison. 
The methods of these studies should also produce reasonably accurate 
measurements of the IMFs in each region (see Luhman et al.\ 2000).

The IMFs for Taurus and the Trapezium are plotted together in 
Figure~\ref{fig:imf}, where both mass functions are representative down to
0.02~$M_{\odot}$. The Trapezium IMF has been normalized to the Taurus
data between 0.1 and 1~$M_{\odot}$. Over this mass range, the two mass 
functions are consistent with the same shape; a peak at 0.6-1~$M_{\odot}$ 
followed by slight decline and flattening down to 0.1~$M_{\odot}$. 
However, below 0.1~$M_{\odot}$ the two IMFs differ greatly.
If the IMF in Taurus were the same as that in the Trapezium, $12.8\pm1.8$ brown 
dwarfs ($>0.02$~$M_{\odot}$) are expected in the former, but only one is found.
Relative to the Trapezium, Taurus exhibits a deficit of higher mass stars 
as well. At 1-2~$M_{\odot}$ the renormalized Trapezium IMF contains 
$7.8\pm1.4$ stars while the Taurus sample has only 2 stars. 
Neither IMF is characterized by a log-normal distribution. 

The deficit of brown dwarfs in Taurus relative to the Trapezium motivates
searches for variations in substellar mass functions in regions of intermediate
stellar densities. For instance, the number statistics of the substellar 
census in IC~348 (Luhman et al.\ 1998b; Luhman 1999) and $\rho$~Oph (Luhman 
\& Rieke 1999; Wilking et al.\ 1999) can be improved to determine 
whether they differ from the Trapezium at a significant level.

The significant number of brown dwarfs in the field (Reid et al.\ 1999) and the
scarcity of them in Taurus combined with the similarity of the IMFs in 
star forming clusters, young open clusters, and the field 
(Luhman et al.\ 2000) are 
convincing evidence that the field is populated by stars born in clusters 
rather than isolated regions (Lada, Strom, \& Myers 1993).

\subsubsection{Implications for Theories of the IMF}

Luhman et al.\ (2000) compared predictions of theories of the IMF 
to the shape of the low-mass IMF, its approximate invariance at 
stellar masses among IC~348, $\rho$~Oph, and the Trapezium, and the minimum mass
observed for isolated objects ($\sim0.01$~$M_{\odot}$), possibly below the
deuterium burning mass limit (0.013-0.015~$M_{\odot}$, Burrows et al.\ 1997).
In addition, while similarities between the mass functions of pre-stellar 
clumps and the stellar IMF have recently been reported and suggested as 
evidence that the process of fragmentation determines the masses of stars 
(Motte, Andr\'{e}, \& Neri 1998; Testi \& Sargent 1998), Luhman et al.\ (2000) 
found that a closer comparison of these pre-stellar mass functions to accurate 
IMFs recently derived for star forming regions is not yet conclusive. 
I now include the implications of the new IMF for 
Taurus and its behavior relative to the mass function for the Trapezium.

The high-mass IMF in Taurus is first considered. 
Elmegreen (1997, 1999, 2000) contended that differences in measurements of
IMF slopes may be statistical fluctuations 
rather than true variations in the IMF. Regions with fewer high 
mass stars simply do not have enough members to populate the mass 
function to high masses. He points to the agreement between cluster IMFs 
and the integrated galaxy IMF as supporting evidence. However, it is clear 
from the work presented here and by Luhman et al.\ (2000) that
star forming clusters rather than sparse regions like Taurus are 
responsible for the Galactic field population. Thus, a true deficit of 
high-mass stars in Taurus would not be reflected in the Galactic IMF. 
Indeed, a comparison of Taurus and the Trapezium suggests that the lack of 
high-mass stars is not because of poor sampling of the IMF, although the
statistical significance of this result should be improved through an 
expansion of this work to include a larger fraction of the Taurus 
population. Elmegreen does refer to possible
exceptions to his model, such as the extreme field in the LMC and Milky
Way (Massey et al.\ 1995), where the high-mass IMFs are much steeper
than those in the
Galactic and cluster IMFs. In fact, Elmegreen (1999) briefly refers to a
scenario in which a low-density star forming region like Taurus could 
produce fewer high-mass stars.

The predictions of individual theories are now compared to the low-mass 
IMFs in Taurus and the Trapezium.

While wind-limited accretion models for the IMF (Adams \& Fatuzzo 1996)
can account for the value of the low-mass cutoff in Taurus,
they do not explain the change in
the minimum mass from Taurus to the Trapezium, the very low value
of the cutoff in the Trapezium, or the deficit of 
high-mass stars in Taurus. Furthermore, if the IMF is determined by a 
large number of variables, a log-normal form is predicted for the IMF,
which is not observed in either region.

Elmegreen (1997, 1999) presented a model where the IMF is controlled by 
fragmentation from hierarchical clouds modulated by the Jeans condition.
The shape of the IMF at intermediate masses in Taurus and the 
Trapezium -- a power law slope followed by a turnover and flattening --
is reproduced by this model (and many others; Scalo 1999). 
Both the turnover mass and the minimum mass should be reflections of the 
Jeans mass, thus they should change together from one region to another.  
However, this is not the case in the Taurus and 
the Trapezium populations, which exhibit the same turnover masses
of $\sim0.8$~$M_{\odot}$ but very different low-mass cutoffs of
$\lesssim0.01$~$M_{\odot}$ and $\sim0.08$~$M_{\odot}$, respectively.

Theories of the IMF that rely on interactions of
cores or protostars (Lejeune \& Bastien 1986; Murray \& Lin 1996; 
Price \& Podsiadlowski 1995; Bonnell et al.\ 1997) have difficulty in 
explaining the similarity of the turnover masses in the Trapezium and Taurus, 
where the stellar densities differ by three orders of magnitude. 
Dynamical interactions and competitive accretion between protostellar cores
were also important in shaping the IMF in numerical modeling of the 
gravitational collapse and fragmentation of a dense cloud core by
Klessen, Burkert, \& Matthew (1998).
The predicted IMF exhibited a log-normal form that was centered at a
mass described by the product of the Jeans mass of the system and the star
forming efficiency. 
As demonstrated here and by Luhman et al.\ (2000), the IMF is not characterized
by a log-normal function, with the possible exception of that in globular 
clusters. 


Any theory of the IMF must explain the differences in the
frequencies of brown dwarfs and (more tentatively) of high-mass stars between
Taurus and young clusters. 
One property that varies dramatically from Taurus
to star forming clusters is the level of turbulence observed in cloud cores
(Myers 1998), which may be the ultimate origin of these IMF variations.
For instance, it is possible
that both the growth and the fragmentation of pre-stellar cores are enhanced
in more turbulent environments, which could broaden the IMF in a dense cluster
relative to that in an isolated region (Myers 2000b).
Further modeling of the formation of stellar clusters within turbulent dense
cores (Klessen et al.\ 1998; Myers 1998, 2000a) may help explain the observed
behavior of the low-mass IMF.

\section{Conclusion}

An IMF that is representative for masses down to 0.02~$M_{\odot}$ has been 
measured for 
fields near the L1495, B209, L1529, and L1551 dark clouds in the Taurus star 
forming region. These data and similar studies of star forming clusters 
provide powerful constraints on theories of star formation. 

The observations for star forming clusters as summarized by Luhman 
et al.\ (2000) are first reviewed:

\begin{enumerate}

\item
Above 0.1~$M_{\odot}$, the IMFs in the core of IC~348, the cloud core of 
$\rho$~Oph, and the Trapezium Cluster are indistinguishable within the
counting uncertainties. 

\item
Below 0.1~$M_{\odot}$, the IMFs in the three clusters 
are roughly similar, although the statistical uncertainties in the 
data for $\rho$~Oph and IC~348 allow for modest variations.

\item
Data for the Trapezium, young open clusters, and the field
are consistent with the same IMF, which differs from
the IMF that characterizes globular clusters (Paresce \& De Marchi 2000).

\item 
The IMF for young clusters and the field is flat or slowly rising 
from the substellar regime to 
$\sim0.6$~$M_{\odot}$ and then rolls over into a power law that
continues from $\sim1$~$M_{\odot}$ to higher masses with a slope similar to
or somewhat larger than the Salpeter value of 1.35; it cannot be
described by a log-normal function.

\item
In the Trapezium, the IMF is flat down to 0.02~$M_{\odot}$ or lower and
contains a population of $\sim50$ likely brown dwarfs. 

\item 
The least massive objects observed in the Trapezium appear to have masses of 
$\sim0.01$~$M_{\odot}$.

\end{enumerate}

The new conclusions provided by this study of Taurus are as follows:

\begin{enumerate}

\item
Above 1~$M_{\odot}$, the fields of Taurus in this study have proportionately 
fewer stars than the Trapezium at a modest level of significance. 

\item
From 0.1-1~$M_{\odot}$, the IMF in Taurus matches that of the Trapezium; 
both regions have a turnover mass near $\sim0.8$~$M_{\odot}$ and a 
slow decline and flattening to lower masses.

\item
Below 0.1~$M_{\odot}$, there is a significant deficit of objects
in Taurus relative to the Trapezium. If the Trapezium IMF is normalized 
to the Taurus IMF by the numbers of objects between 0.1-1~$M_{\odot}$, 
then $12.8\pm1.8$ brown dwarfs at $>0.02$~$M_{\odot}$ are present in the 
Trapezium where one is found in the Taurus fields.

\item
In summary, if the Trapezium IMF is steepened above 1~$M_{\odot}$ and
suppressed below 0.1~$M_{\odot}$, then the result is the IMF 
for Taurus; it cannot be described by a log-normal function.

\end{enumerate}

\acknowledgements
I am grateful to F. Allard, I. Baraffe, and F. D'Antona for access to 
their most recent calculations.  Discussions with L. Hartmann,
G. Rieke, J. Stauffer and, in particular, P. Myers are appreciated.  
I also thank C. Brice\~{n}o for 
providing details concerning his published work.  K. L. was funded by a 
postdoctoral fellowship at the Harvard-Smithsonian Center for Astrophysics.  
This publication makes use of data products from the Two Micron All Sky Survey, 
which is a joint project of the University of Massachusetts and the
Infrared Processing and Analysis Center, funded by the National Aeronautics and 
Space Administration and the National Science Foundation.

\newpage

\clearpage
 
\clearpage
 

\clearpage
 
\clearpage
 
\clearpage
 
\clearpage
 
\end{document}


\pagestyle{empty}
\begin{deluxetable}{llllllllllllllll}
\scriptsize
\tablewidth{0pt}
\tablecaption{Young Stars in Selected Fields of Taurus}
\tablehead{\colhead{ID\tablenotemark{a}}
& \colhead{$\alpha$(2000)} & \colhead{$\delta$(2000)}
& \colhead{Optical Type\tablenotemark{b}} & \colhead{IR Type\tablenotemark{b}}
& \colhead{Adopt} & \colhead{$T_{\rm eff}$\tablenotemark{c}}
& \colhead{$A_H$} & \colhead{$L_{\rm bol}$} & \colhead{$R-I$}
& \colhead{$I-z\arcmin$} & \colhead{$I$}
& \colhead{$J-H$} & \colhead{$H-K_s$}
& \colhead{$K_s$}
& \colhead{in IMF?} \\
}
\startdata
                      Anon 1 &  4\ 13\ 27.21 & 28\ 16\ 24.9 &            M0(1) &         \nodata &         M0 &   3850 &    0.64 &     2.6 &      \nodata &     \nodata &       \nodata &     1.07 &     0.36 &     7.41 &  yes \nl
                    V773 Tau &  4\ 14\ 12.91 & 28\ 12\ 12.4 &            K3(1) &         \nodata &         K3 &   4730 &    0.29 &     6.5 &    0.85 &     \nodata &       \nodata &     0.85 &     0.43 &     6.21 &  yes \nl
                      FM Tau &  4\ 14\ 13.57 & 28\ 12\ 49.3 &            M0(1) &         \nodata &         M0 &   3850 &    0.34 &    0.44 &    1.16 &     \nodata &       \nodata &     0.86 &     0.70 &     8.75 &  yes \nl
                      CW Tau &  4\ 14\ 17.00 & 28\ 10\ 57.8 &            K3(1) &         \nodata &         K3 &   4730 &    0.34 &     1.1 &    0.91 &     \nodata &       \nodata &     1.30 &     1.12 &     7.12 &  yes \nl
                     CIDA-1 &  4\ 14\ 17.61 & 28\ 06\ 09.8 &             C(1) &         \nodata &         \nodata &     \nodata &      \nodata &      \nodata &      \nodata &   1.42 &    14.33 &     1.13 &     0.68 &     9.89 &   no \nl
                       MHO 1 &  4\ 14\ 26.27 & 28\ 06\ 03.2 &           M2.5(3) &         \nodata &       M2.5 &   3488 & $\sim1.0$ & $\sim0.14$ &    2.40 &   2.12 &    17.84 &     2.49 &     2.21 &     7.71 &  yes \nl
                       MHO 2 &  4\ 14\ 26.40 & 28\ 05\ 59.7 &           M2.5(3) &         \nodata &       M2.5 &   3488 &    1.15 &    0.42 &    2.60 &   1.89 &    15.79 &     2.20 &     1.47 &     7.78 &  yes \nl
                       MHO 3 &  4\ 14\ 30.55 & 28\ 05\ 14.6 &             K7(3) &         \nodata &         K7 &   4060 &    1.33 &    0.91 &    2.36 &   1.64 &    14.49 &     1.93 &     1.02 &     8.21 &  yes \nl
                      FO Tau &  4\ 14\ 49.28 & 28\ 12\ 30.4 &     M2(1),M2(2) &         \nodata &         M2 &   3560 &    0.61 &     1.1 &      \nodata &     \nodata &       \nodata &     1.07 &     0.48 &     8.11 &  yes \nl
                     CIDA-2 &  4\ 15\ 05.16 & 28\ 08\ 46.1 &          M4.5(1) &         \nodata &       M4.5 &   3198 &    0.10 &    0.28 &      \nodata &     \nodata &       \nodata &     0.68 &     0.34 &     9.08 &  yes \nl
                     V410 X-ray 1 &  4\ 17\ 49.65 & 28\ 29\ 36.4 &            M4(2) &         \nodata &         M4 &   3270 &    0.16 &    0.14 &    1.79 &   1.30 &    13.78 &     1.30 &     0.62 &     9.08 &  yes \nl
                     V410 X-ray 3 &  4\ 18\ 07.98 & 28\ 26\ 03.7 &     M6(2,5,3) & M4.5-M8.5(4) &         M6 &   2990 &    0.14 &   0.089 &    2.27 &   1.57 &    14.18 &     0.71 &     0.39 &    10.39 &  yes \nl
                     V410 Anon 13 &  4\ 18\ 17.11 & 28\ 28\ 41.8 &   M5(2),M5-M7(3) & K2-K6(4) &         M5 &   3125 &    0.67 &   0.048 &    2.61 &   1.88 &    16.59 &     1.24 &     0.71 &    10.95 &  yes \nl
                     V410 Anon 24 &  4\ 18\ 22.41 & 28\ 24\ 37.3 &                \nodata & F9-G3(4) &         G1 &   5945 &    4.18 &     2.8 &      \nodata &   2.50 &    21.73 &     2.85 &     1.57 &    10.68 &   no \nl
                     V410 Anon 25 &  4\ 18\ 29.09 & 28\ 26\ 19.1 &                \nodata & K7-M3(4) &         M1 &   3705 &    4.06 &     1.6 &      \nodata &     \nodata &       \nodata &     3.22 &     1.76 &     9.87 &   no \nl
                    V410 Tau &  4\ 18\ 31.10 & 28\ 27\ 16.1 &     K3(1),K7(2) &         \nodata &         K7 &   4060 &    0.00 &     1.5 &    0.70 &     \nodata &       \nodata &     0.65 &     0.17 &     7.61 &  yes \nl
                      DD Tau &  4\ 18\ 31.13 & 28\ 16\ 28.9 &     M1(1),M4(2) &         \nodata &         M4 &   3270 &    0.00 &    0.32 &    1.44 &     \nodata &       \nodata &     1.13 &     0.81 &     7.87 &  yes \nl
                      CZ Tau &  4\ 18\ 31.58 & 28\ 16\ 58.5 &   M1.5(1),M3(2) &         \nodata &         M3 &   3415 &    0.32 &    0.28 &    1.73 &     \nodata &       \nodata &     0.72 &     0.45 &     9.34 &  yes \nl
              PSC 04154+2823 &  4\ 18\ 32.05 & 28\ 31\ 15.3 &                \nodata & M0.5-M4.5(4) &       M2.5 &   3488 & $\sim2.6$ & $\sim0.13$ &      \nodata &   2.14 &    19.65 &     2.72 &     2.15 &    10.25 &   no \nl
                     V410 X-ray 2 &  4\ 18\ 34.44 & 28\ 30\ 30.2 &                \nodata & K6-M2(4) &         M0 &   3850 &    3.74 &     3.0 &      \nodata &     \nodata &       \nodata &     2.99 &     1.55 &     9.16 &   no \nl
                     V410 X-ray 4 &  4\ 18\ 40.23 & 28\ 24\ 24.4 &                \nodata & M3-M5(4) &         M4 &   3270 &    3.28 &     1.3 &      \nodata &   2.48 &    19.91 &     2.62 &     1.29 &     9.64 &   no \nl
                    V892 Tau &  4\ 18\ 40.61 & 28\ 19\ 15.4 &     A6(1),B9(2) &         \nodata &         B9 &  10500 &    1.48 &      77 &    1.71 &     \nodata &       \nodata &     1.71 &     1.24 &     5.76 &   no \nl
                        LR 1 &  4\ 18\ 41.33 & 28\ 27\ 24.9 &                \nodata &   K4-K5(8) &       K4.5 &   4470 &    3.97 &    0.47 &      \nodata &     \nodata &       \nodata &     3.21 &     1.94 &    10.98 &   no \nl
                     V410 X-ray 7 &  4\ 18\ 42.49 & 28\ 18\ 49.7 &    M1(2),M0.5(3) &         \nodata &      M0.75 &   3741 &    1.45 &    0.50 &    2.62 &   1.80 &    15.92 &     1.83 &     0.85 &     9.23 &   no \nl
                     V410 Anon 20 &  4\ 18\ 45.05 & 28\ 20\ 52.9 &                \nodata & K1-K5(4) &         K3 &   4730 &    4.08 &    0.51 &      \nodata &     \nodata &       \nodata &     2.99 &     1.46 &    11.92 &   no \nl
                    Hubble 4 &  4\ 18\ 47.03 & 28\ 20\ 07.3 &     K7(1),K7(2) &         \nodata &         K7 &   4060 &    0.42 &     2.7 &    1.27 &     \nodata &       \nodata &     0.92 &     0.32 &     7.28 &  yes \nl
                  CoKu Tau/1 &  4\ 18\ 51.48 & 28\ 20\ 26.5 &      M0(7,M2(2) &         \nodata &         M0 &   3850 &    0.68 &   0.068 &    1.57 &   1.50 &    15.96 &     1.38 &     0.50 &    11.01 &  yes \nl
              PSC 04158+2805 &  4\ 18\ 58.14 & 28\ 12\ 23.7 &                \nodata & K7-M3(4) &         M1 &   3705 &    1.02 &   0.047 &    2.14 &   1.55 &    16.43 &     1.39 &     1.21 &    11.17 &  yes \nl
                     V410 X-ray 6 &  4\ 19\ 01.10 & 28\ 19\ 42.1 &    M5(2),M5.5(6) & M2.5-M6.5(4) &       M5.5 &   3058 &    0.11 &    0.20 &    2.08 &   1.59 &    13.48 &     1.00 &     0.34 &     9.18 &  yes \nl
                    V410 X-ray 5a &  4\ 19\ 01.98 & 28\ 22\ 33.3 &  M5(2),M5.5(6,3) & M3-M5(4) &       M5.5 &   3058 &    0.45 &   0.083 &    2.49 &   1.73 &    15.37 &     1.24 &     0.61 &    10.14 &  yes \nl
                      FQ Tau &  4\ 19\ 12.81 & 28\ 29\ 33.1 &     M2(1),M4(2) &         \nodata &         M4 &   3270 &    0.10 &    0.21 &    1.72 &     \nodata &       \nodata &     0.79 &     0.36 &     9.30 &  yes \nl
                    V819 Tau &  4\ 19\ 26.26 & 28\ 26\ 14.3 &     K7(1),K7(2) &         \nodata &         K7 &   4060 &    0.30 &    0.91 &    1.12 &     \nodata &       \nodata &     0.86 &     0.23 &     8.40 &  yes \nl
                       MHO 9 &  4\ 31\ 15.94 & 18\ 20\ 04.7 &             M4(3) &         \nodata &         M4 &   3270 &    0.39 &    0.24 &    2.07 &   1.27 &    12.95 &       \nodata &       \nodata &       \nodata &  yes \nl
                       MHO 4 &  4\ 31\ 24.08 & 18\ 00\ 21.6 &        M6-M6.5(3) &         \nodata &      M6.25 &   2965 &    0.17 &   0.059 &    2.36 &   1.61 &    14.32 &       \nodata &       \nodata &       \nodata &  yes \nl
                 L1551/IRS5 &   4\ 31\ 34.1 &   18\ 08\ 04 &           G-K(1) &         \nodata &         \nodata &     \nodata &      \nodata &      \nodata &      \nodata &     \nodata &       \nodata &     2.08 &     1.74 &     9.27 &   no \nl
                    LkH$\alpha$ 358 &  4\ 31\ 36.27 & 18\ 13\ 40.9 &          M5.5(1) &         \nodata &       M5.5 &   3058 &    2.38 &    0.59 &      \nodata &   1.49 &    16.05 &     2.09 &     1.21 &     9.69 &   no \nl
                      HL Tau &   4\ 31\ 38.4 &   18\ 13\ 59 &            K7(1) &         \nodata &         M0 &   3850 & $\sim0.3$ & $\sim0.41$ &    1.11 &     \nodata &       \nodata &     1.36 &     1.61 &     7.34 &  yes \nl
                      XZ Tau &   4\ 31\ 40.0 &   18\ 13\ 58 &            M3(1) &         \nodata &         M3 &   3415 &    0.05 &    0.32 &    1.41 &     \nodata &       \nodata &     0.95 &     0.91 &     8.05 &  yes \nl
                      HK Tau &  4\ 31\ 50.55 & 24\ 24\ 18.6 &          M0.5(1) &         \nodata &       M0.5 &   3778 &    0.60 &    0.56 &    1.56 &     \nodata &       \nodata &     1.18 &     0.67 &     8.59 &  yes \nl
                   V710 Tau A &   4\ 31\ 57.7 &   18\ 21\ 37 &            M1(1) &         \nodata &         M1 &   3705 &    0.00 &    0.39 &      \nodata &     \nodata &       \nodata &     0.76 &     0.37 &     8.69 &  yes \nl
                   V710 Tau B &   4\ 31\ 57.7 &   18\ 21\ 37 &            M3(1) &         \nodata &         M3 &   3415 &    0.47 &    0.46 &      \nodata &     \nodata &       \nodata &     0.97 &     0.41 &     8.82 &  yes \nl
                   L1551-51 &  4\ 32\ 09.29 & 17\ 57\ 23.1 &            K7(1) &         \nodata &         K7 &   4060 &    0.00 &    0.47 &    0.70 &     \nodata &       \nodata &     0.67 &     0.19 &     8.85 &  yes \nl
                    V827 Tau &   4\ 32\ 14.5 &   18\ 20\ 15 &            K7(1) &         \nodata &         K8 &   3955 &    0.20 &     1.0 &    0.99 &     \nodata &       \nodata &     0.76 &     0.21 &     8.19 &  yes \nl
                   Haro 6-13 &  4\ 32\ 15.43 & 24\ 29\ 00.0 &             C(1) &         \nodata &         \nodata &     \nodata &      \nodata &      \nodata &      \nodata &   1.46 &    13.54 &     1.89 &     1.24 &     8.09 &   no \nl
                    V826 Tau &   4\ 32\ 15.8 &   18\ 01\ 39 &            K7(1) &         \nodata &         K7 &   4060 &    0.12 &    0.95 &    0.90 &     \nodata &       \nodata &     0.75 &     0.15 &     8.25 &  yes \nl
                       MHO 5 &  4\ 32\ 16.02 & 18\ 12\ 43.4 &        M6-M6.5(3) &         \nodata &      M6.25 &   2965 &    0.02 &   0.065 &    2.18 &   1.61 &    13.72 &       \nodata &       \nodata &       \nodata &  yes \nl
                    V928 Tau &  4\ 32\ 18.88 & 24\ 22\ 27.4 &          M0.5(1) &         \nodata &       M0.5 &   3778 &    0.45 &     1.1 &    1.38 &     \nodata &       \nodata &     1.07 &     0.35 &     8.09 &  yes \nl
                       MHO 6 &  4\ 32\ 22.10 & 18\ 27\ 35.7 &             M5(3) &         \nodata &         M5 &   3125 &    0.15 &   0.062 &    1.98 &   1.34 &    13.80 &       \nodata &       \nodata &       \nodata &  yes \nl
                       MHO 7 &  4\ 32\ 26.27 & 18\ 27\ 45.2 &             M5(3) &         \nodata &         M5 &   3125 &    0.07 &   0.087 &    1.89 &   1.35 &    13.18 &       \nodata &       \nodata &       \nodata &  yes \nl
                      FY Tau &  4\ 32\ 30.59 & 24\ 19\ 57.3 &            K7(1) &         \nodata &         K7 &   4060 &    0.95 &     1.6 &      \nodata &     \nodata &       \nodata &     1.31 &     0.64 &     8.01 &  yes \nl
                      FZ Tau &  4\ 32\ 31.78 & 24\ 20\ 03.0 &             C(1) &         \nodata &         \nodata &     \nodata &      \nodata &      \nodata &      \nodata &     \nodata &       \nodata &     1.51 &     1.06 &     7.31 &   no \nl
                       MHO 8 &  4\ 33\ 01.99 & 24\ 21\ 00.1 &           M5.5(3) &         \nodata &       M5.5 &   3058 &    0.11 &    0.14 &    2.08 &   1.61 &    13.63 &     0.74 &     0.38 &     9.74 &  yes \nl
                    V830 Tau &  4\ 33\ 10.04 & 24\ 33\ 43.4 &            K7(1) &         \nodata &         K7 &   4060 &    0.08 &    0.78 &    0.86 &     \nodata &       \nodata &     0.72 &     0.18 &     8.42 &  yes \nl
\enddata
\tablenotetext{a}{
Designations are from Brice\~{n}o et al.\ 1998 (MHO), Luhman \& Rieke 1998
(LR), Strom \& Strom 1994 (V410~X-ray, V410~Anon), and references within
Kenyon \& Hartmann 1995.}
\tablenotetext{b}{
1=Kenyon \& Hartmann 1995 and references therein, 2=Strom \& Strom 1994, 
3=Brice\~{n}o et al.\ 1998, 4=Luhman \& Rieke 1998, 5=Luhman et al.\ 1998a, 
6=Luhman 1999, 7=Kenyon et al.\ 1998, 8=this work.}
\tablenotetext{c}{Intermediate temperature scale of Luhman 1999.}
\tablecomments{$JHK_s$ data are from the 2MASS survey when available. For
L1551/IRS5, V826~Tau, XZ~Tau, HL~Tau, V827~Tau, V710~Tau~A, V710~Tau~B, and 
LkH$\alpha$~358, $JHK$ data are from Kenyon \& Hartmann 1995. $R-I$ data
are from Brice\~{n}o et al.\ 1998, Strom \& Strom 1994, and
Kenyon \& Hartmann 1995. $I$ and $I-z\arcmin$ are from this work. 
Most coordinates
are from the 2MASS survey or are measured from the $I$ images, where the plate 
solutions were derived using 2MASS sources. The coordinates listed with lower
precision are taken from Herbig \& Bell 1988. Sources without optical data
are saturated or not detected. Sources without IR data are within a portion
of the L1551 field that is not currently available through 2MASS.}
\end{deluxetable}